# Integrated probabilistic computer using voltage-controlled magnetic tunnel junctions as its entropy source

Christian Duffee[1,†], Jordan Athas[1,†], Yixin Shao[1], Noraica Davila Melendez[2], Eleonora Raimondo[3,4], Jordan A. Katine[2], Kerem Y. Camsari[5], Giovanni Finocchio[4], Pedram Khalili Amiri[1,*]

[1]Department of Electrical and Computer Engineering, Northwestern University, Evanston, IL 60208, United States of America
[2]Western Digital Corporation, San Jose, CA 95119, United States of America
[3]Istituto Nazionale di Geofisica e Vulcanologia, Rome 00143, Italy
[4]Department of Mathematical and Computer Sciences, Physical Sciences and Earth Sciences, University of Messina, Messina 98166, Italy
[5]Department of Electrical and Computer Engineering, University of California, Santa Barbara, CA 93106, United States of America
[*]Corresponding author: Pedram Khalili Amiri (pedram@northwestern.edu)
[†]These authors contributed equally to this work.

## ABSTRACT

Probabilistic Ising machines (PIMs) provide a path to solving many computationally hard problems more efficiently than deterministic algorithms on von Neumann computers. Stochastic magnetic tunnel junctions (S-MTJs), which are engineered to be thermally unstable, show promise as entropy sources in PIMs. However, scaling up S-MTJ-PIMs is challenging, as it requires fine control of a small magnetic energy barrier across large numbers of devices. In addition, non-spintronic components of S-MTJ-PIMs to date have been primarily realized using general-purpose processors or field-programmable gate arrays. Reaching the ultimate performance of spintronic PIMs, however, requires co-designed application-specific integrated circuits (ASICs), combining CMOS with spintronic entropy sources. Here we demonstrate an ASIC in 130 nm foundry CMOS, which implements integer factorization as a representative hard optimization problem, using PIM-based invertible logic gates realized with 1143 probabilistic bits. The ASIC uses stochastic bit sequences read from an adjacent voltage-controlled (V-) MTJ chip. The V-MTJs are designed to be thermally stable in the absence of voltage, and generate random bits on-demand in response to 10 ns pulses using the voltage-controlled magnetic anisotropy effect. We experimentally demonstrate the chip's functionality and provide projections for designs in advanced nodes, illustrating a path to millions of probabilistic bits on a single CMOS+V-MTJ chip.

## I. Introduction

Physics-inspired unconventional computing methodologies offer promise in addressing hard computing problems - particularly those classified as non-deterministic polynomial hard (NP-hard) - more efficiently than approaches based on deterministic algorithms implemented on von Neumann architectures[1]. For many problems in this category, even modest one-time improvements in solution speed or solution quality yield significant practical and economic impacts. Examples include the optimization of transportation networks, logistics, and energy grids, all of which bear major sustainability implications. Additionally, advancements in this realm extend to cryptography through the solution of large integer factorization problems[2–6].

An important family of unconventional computing methods is based on Ising machines[7–9], where computational problems are mapped onto a network of interacting and energy-tunable bistable elements. These elements (i.e., Ising "spins") are akin to the spins of the original Ising model in magnetism[10,11]. They collectively evolve towards the global energy minimum (or minima) of the network, which are designed to correspond to the solution(s) of the selected computing problem.

Various physical platforms can be used to implement Ising machines. Examples include networks of optical parametric oscillators[12–17], complementary metal-oxide semiconductor (CMOS) oscillators[18–23], simulated bifurcation machines[24–26], spin wave oscillators[27–29], probabilistic (p-) bits made from stochastic magnetic tunnel junctions (S-MTJs)[30–34], and superconducting quantum devices[35,36]. Combinatorial optimization problems can be often expressed in terms of an effective number of bits. However, current implementations of Ising machines are constrained to relatively small problem sizes, due to the scaling challenges inherent to each physical platform. As an example, MTJ-based Ising machines, combined with conventional CPUs or field-programmable gate arrays (FPGAs), have been used to perform integer factorization of numbers up to 40 bits in length[37,38]. However, significantly larger numbers of bits are often required to be of practical interest[31,40].

Here we propose hybrid CMOS-spintronic application-specific integrated circuits (ASICs) as a pathway towards scalable probabilistic Ising machines (PIMs). In comparison to FPGAs and microcontrollers, which have also been used to implement PIMs[6,9,34,41], an optimized ASIC architecture can achieve higher density and enhanced performance at a comparable node size[42]. Our proposed hardware therefore combines the established capabilities of digital CMOS with ultrafast and compact spintronic entropy sources, which are difficult to implement efficiently in existing CMOS-only solutions.

While this approach to realizing PIMs can be used for a wide range of applications, here we illustrate its potential by focusing on integer factorization as a representative hard optimization problem, and demonstrate a factorization PIM using a custom-designed ASIC in a 130 nm foundry process. Our ASIC contains nearly all necessary components for integer factorization in a p-computing scheme, including the implementation of p-bits and their weighted connections which define the PIM. Additionally, we designed a true random number generator (TRNG) using voltage-controlled MTJs (V-MTJs) addressed with an access printed circuit board (PCB) as the entropy source. The voltage-controlled magnetic anisotropy (VCMA) effect is used to reliably generate an on-demand random bit from a V-MTJ with high speed (~ 10 ns/bit) and energy efficiency (< 0.43 pJ) [37,43]. The surrounding control circuitry prepares and records the generated random bit, which is then queried and used by the ASIC in the PIM implementation. The implementation also follows a synchronous design to reduce the issue of passive power consumption encountered in asynchronous hardware[33]. The choice of VCMA-based TRNGs is driven by their out-of-plane magnetic configuration and large energy barrier in the absence of voltage, which make them more similar to the MTJs adopted by foundries for memory production. Importantly, unlike S-MTJs, on-demand random bit generation with VCMA does not require a small, fine-tuned energy barrier which can be difficult to uniformly realize in large arrays. V-MTJs also have decreased power consumption and access transistor size compared to conventional current-controlled MTJ devices[44].

We experimentally demonstrate the capability of our combined ASIC and V-MTJ system to solve 6-bit integer factorization problems. While the size of problems that can be solved experimentally by the present chip is limited by the slow input and output (I/O) speed between the ASIC and the V-MTJs, we also show solution of a 20-bit factorization problem using a slightly modified simulated design. Additionally, we adapt the ASIC design to the 45 nm and 7 nm process nodes, to evaluate the scalability of this approach in solving large-scale p-computing problems that require millions of p-bits integrated on-chip.

## II. Overview of PIM Architecture

In a probabilistic Ising machine, a collection of Ising spins (p-bits) stochastically flips between states to explore an energy landscape whose minima represent solutions to a computational problem[1,9,30,31,40,45–50]. PIMs can be implemented via a variety of mechanisms so long as the state of each p-bit ($m_i$) is updated after each time step ($t$) according to

$$m_i(t+1) = sgn\left[tanh\left(\frac{I_i}{T}\right) + r_{uni(-1,1)}\right] \quad (1)$$

based on a random number from a uniform distribution ($r$), and an input ($I_i$) scaled by the temperature ($T$)[9,40,49]. The input itself is given by the gradient of the problem-defining energy landscape. If $J$ is the weight connectivity matrix and $h$ is the bias vector for $N$ p-bits,

$$I_i = -\frac{\partial}{\partial m_i}E = -\frac{\partial}{\partial m_i}\left[-\left(\sum_{i=1}^{N} h_i m_i + \sum_{i=1}^{N}\sum_{j=1}^{N}\frac{1}{2}J_{ij}\,m_i m_j\right)\right] = h_i + \sum_{j=1}^{N} J_{ij}\,m_j. \quad (2)$$

Solving a computing problem in this manner requires two key components: (A) A high-quality source of randomness (i.e., an entropy source), which can generate truly-random, or very high-quality pseudo-random, numbers in vast quantities and with high throughput; and (B) a high-density physical platform capable of implementing the interconnections and bias terms of the PIM (i.e., the $J_{ij}$, $h_i$, and *tanh* terms in Equations 1 and 2).

## III. Design of Voltage-controlled Magnetic Entropy Source

Generating good pseudo-random numbers using traditional CMOS hardware requires extensive resources, making external non-CMOS TRNGs an attractive choice for probabilistic computing hardware[34]. In this work, we use perpendicular V-MTJs[51–53] with a relatively large voltage-dependent energy barrier ($E_b$) between their two free layer states for this purpose[54–57]. A schematic of these devices is shown in Fig. 1a. When voltage is applied across the device, the VCMA effect reduces the value of $E_b$[58,59] and correspondingly, the free layer coercivity, as shown in Fig. 1b. If the duration of the applied voltage pulse is longer than a few nanoseconds, the magnetization orientation of the MTJ's free layer will fall in-plane and precess around the effective field, the orientation of which is determined by a combination of the shape anisotropy and a small in-plane external magnetic field. Upon voltage removal, the large $E_b$ gets reinstated, causing the magnetization to re-enter one of the two perpendicular states at random. This procedure, which is illustrated in Fig. 2, allows for electrically triggered fast (~ 10 ns/bit per V-MTJ) on-demand random bit generation[37,60]. Furthermore, this structure can be implemented within a single-transistor cell similar to existing magnetic random-access memory (MRAM) arrays, providing the possibility of reliable manufacturing of large V-MTJ numbers[61,62].

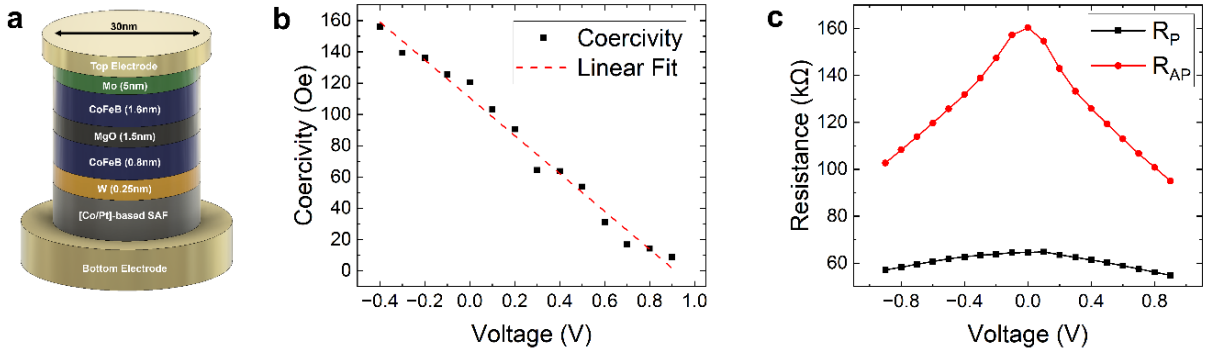

**Figure 1. V-MTJ characteristics. a** Schematic illustration of the 30 nm V-MTJ pillar used in this work. Voltage is applied between the top and bottom electrodes. **b** Relation between applied voltage and coercivity of the device due to VCMA. A linear interpolation is accurate over the voltage range explored in this work. **c** Relation between the read voltage and the parallel and antiparallel resistance levels of the V-MTJ.

It is worth comparing this approach with the more conventional method of using S-MTJs with a small energy barrier to generate random numbers[33,34]. In the latter approach, a single MTJ can, in principle, implement both the randomness and nonlinearity terms of Equation 1. However, this comes at the price of having to choose a very small (typically < 5 kT) steady-state energy barrier, which results in large device-to-device variations. By contrast, the VCMA-induced random bit generation approach can achieve similar bit rates, while allowing for large steady-state energy barriers (> 30 kT in the absence of voltage in this work).

A V-MTJ material stack composed of a Co/Pt-based synthetic antiferromagnetic (SAF)[63,64] layer / W (0.25 nm) / $Co_{20}Fe_{60}B_{20}$ (0.8 nm) / MgO (1.5 nm) / $Co_{17.5}Fe_{52.5}B_{30}$ (1.6 nm) / Mo (5 nm) was used due to its previously demonstrated large VCMA coefficient[60]. The devices had a pillar diameter of 30 nm and are schematically illustrated in Fig. 1a.

We initially performed individual measurements of the V-MTJ devices using a ground-signal (GS) probe. These measurements were performed on a probe station using a current-controlled electromagnet with an adjustable field angle. In these measurements, we define voltage applied to the top electrode relative to the bottom electrode as positive. The probe was simultaneously connected to a sourcemeter for readout and a high-speed voltage pulse generator for writing via a bias tee.

Initially, we positioned the V-MTJ such that the applied magnetic field was oriented purely out-of-plane (parallel to the V-MTJ free layer). The device resistance was then recorded over a range of bias voltages, while sweeping the external magnetic field from -1000 Oe to 1000 Oe and back. The variation of coercivity with bias voltage in Fig. 1b confirms the presence of the VCMA effect. The device had a maximum tunneling magnetoresistance (TMR) ratio of approximately 150% (Fig. 1c) at low read voltages, which provides a large enough resistive difference to reliably determine the magnetic state and convert it into a digital voltage using a voltage divider and comparator circuit.

A constant tilted magnetic field was applied at an offset angle of 30° to 40° from the device normal. The modified field direction was necessary to compensate the out-of-plane stray field from the V-MTJ fixed layer, while also providing an in-plane component around which the magnetization could precess after the application of each voltage pulse. To generate a random bit, a single 10 ns write voltage pulse was applied across the device. This lowered the height of the energy barrier between the parallel (P) and antiparallel (AP) free layer orientations, effectively merging them into a single (in-plane) state. After the pulse was removed, the energy barrier was restored, and a resistance measurement was conducted. Fig. 2a depicts how the V-MTJ's energy landscape is dynamically modified throughout this process. In this work, a low resistance, indicating a parallel state, is treated as a "0" and a high resistance, indicating an antiparallel state, is treated as a "1". However, as this is a 50%/50% random number generator, the inverse mapping could also be used. Importantly, unlike deterministic VCMA-induced switching which requires timing the voltage pulse to coincide with half a precession period of the free layer, the mechanism for random bit generation used here only requires pulses to be long enough for the damped magnetization precession to reach its in-plane equilibrium. Typically, this occurs after several full precession periods (10 ns in the present work). As a result, the TRNG is less sensitive to small variations in pulse length.

The minimum voltage required for generation of the random bit inversely depends on the VCMA coefficient. Assuming a constant standby energy barrier, reducing the MTJ diameter would therefore require materials with a larger perpendicular magnetic anisotropy and VCMA coefficient. However, it is worth noting that unlike the case of nonvolatile memory applications where data needs to be preserved for long periods of time, probabilistic applications can accommodate smaller standby energy barrier (and hence VCMA) values. A sample bit stream generated from a V-MTJ in this manner is shown in Fig. 2b[65].

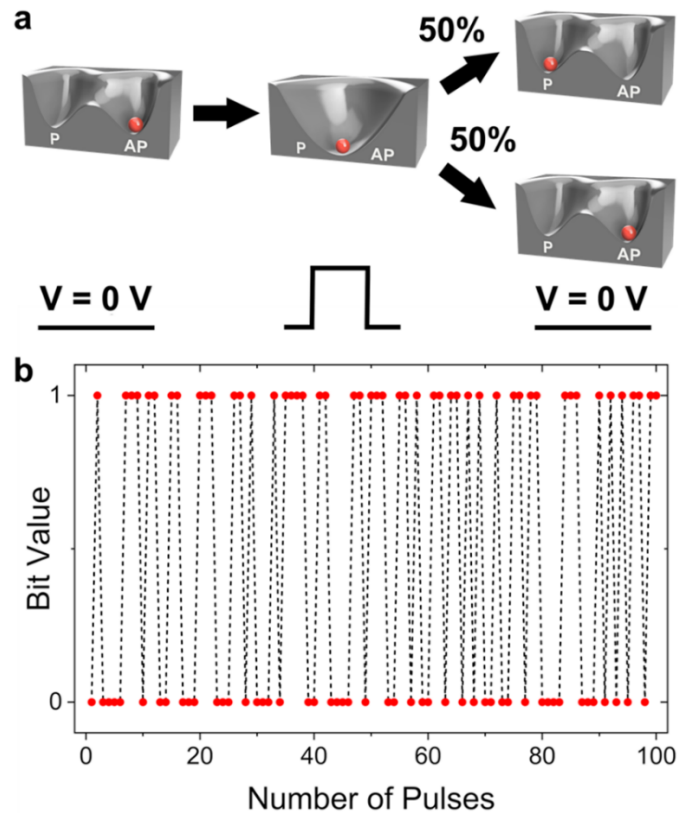

**Figure 2. VCMA-based random bit generation. a** Energy landscape illustration of MTJ undergoing VCMA-induced random switching. When a voltage pulse is applied to the MTJ, the energy barrier between the parallel (P) and antiparallel (AP) states disappears, causing the system to relax to an in-plane intermediate state within a few nanoseconds. When the voltage pulse is removed, the energy barrier re-emerges, and the system finds itself randomly relaxing to either the P or AP state. **b** Measured digital state of a V-MTJ over 100 voltage pulses with 10 ns width. A constant tilted magnetic field of -62 Oe at ~37° from the device normal was applied to achieve proper biasing.

## IV. P-computing System Design

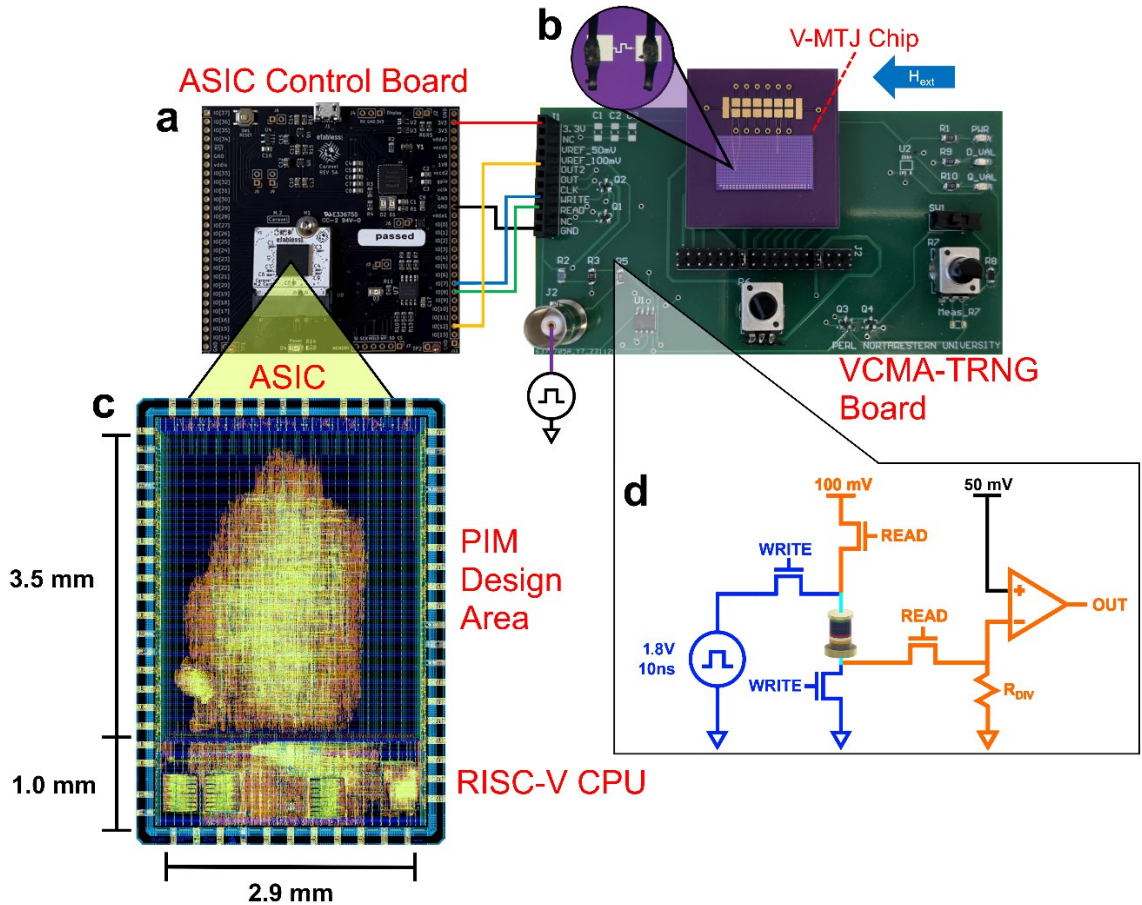

**Figure 3. Experimental configuration. a** Setup showing the connections between the ASIC, the ASIC's control board, the VCMA-TRNG PCB, the V-MTJ, and the pulse generator. **b** Microscope photograph of the connections to a V-MTJ device. **c** The Graphic Design System (GDSII) format rendering of the ASIC design, excluding metal layer 1 for visibility. The design area, center top, is connected to the ASIC control board and the RISC-V CPU, bottom center. As the design is primarily composed of many small p-bit modules, there are no clear large-scale patterns. **d** Diagram of the read and write circuitry for a V-MTJ implemented within the VCMA-TRNG PCB. In read mode, the V-MTJ resistance is compared against a resistor whose value falls between the V-MTJ's P and AP resistances. The write path, the read path, and their overlap are shown in blue, orange, and light blue, respectively.

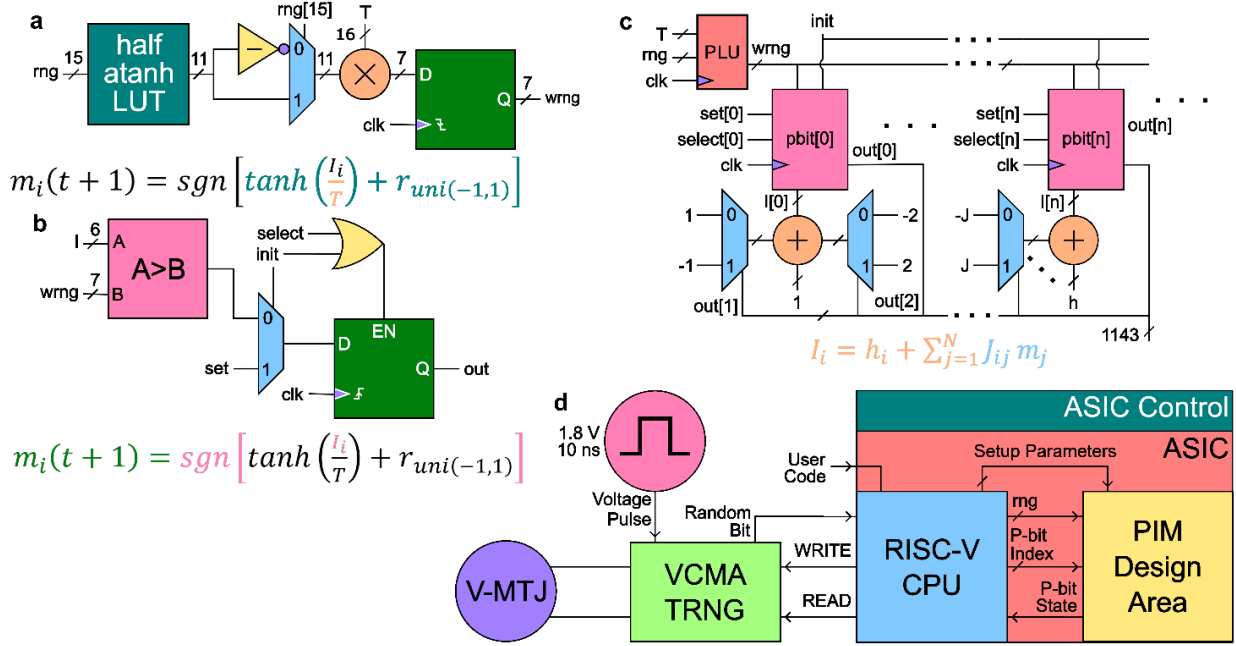

**Figure 4. Data flow block diagram. a** The PLU which converts a uniformly distributed random number to one pulled from the hyperbolic arc-tangent distribution, scaled by a mathematical temperature. This implements part of the probabilistic update equation, with the corresponding portions color-coded to match the circuit. The "rng" signal is a collection of random bits that can be provided externally from V-MTJs. **b** A p-bit which holds a binary state, which is synchronously updated when the p-bit's "select" signal is high, by comparing an input to a PLU-provided sample from a scaled hyperbolic arc-tangent random distribution. This implements the remaining part of the probabilistic update equation. The p-bit's initial state can be set externally with an "init" flag. **c** The ASIC design as a whole, excluding control circuitry. The input equation for each p-bit is calculated here as its $h$ value added to a $J$-vector scaled sum of other p-bit values. **d** The block diagram of the designed system as a whole. Experiments, specified by user code, are run on the RISC-V CPU which initiates the PIM design area's initial state, provides it random bits gathered from the V-MTJ, and reads out the resultant p-bit states with a provided index.

To perform integer factorization, a system was implemented consisting of (i) an ASIC PIM chip, connected to an ASIC control board, and (ii) the V-MTJ-based TRNG PCB. This setup is shown in Fig. 3a. The setup, excluding the pulse generator, ran off a single 5 W USB-A port whose 5 V supply was stepped down by the ASIC control board to the 1.8 V needed by the ASIC and the 3.3 V needed by the other parts of the system. The ASIC PIM and its control board measured a power consumption of 395 mW while problem solving. The TRNG PCB during operation consumes 9 mW of power.

A chip containing 2400 V-MTJs, shown in Fig. 3b and Supplementary Note 1, was wire-bonded to a small PCB which was itself connected to a custom-designed VCMA-TRNG access board via a pin header. A single V-MTJ was used to generate all random numbers for each reported experiment. The wire bonding process did not affect the measured TMR. A fixed magnet, placed above the VCMA-TRNG board, was used to provide the necessary magnetic field for stay field compensation and procession reference. We determined that a single ~1.8 V voltage pulse with 10 ns duration, as shown in Supplementary Note 1, was sufficient for reliably generating random numbers from this 30 nm device. Due to the relatively high resistance of the V-MTJ and the narrowness of the applied voltage pulse, each writing pulse only consumed approximately 0.43 pJ, which is negligible compared to the passive energy consumption of the paired

CMOS circuitry. Reading consumed even less power, requiring approximately 1.3 fJ across the device for a 10 ns read cycle.

The PIM factorization ASIC was designed and fabricated using an open-source 130 nm process design kit (PDK) from SkyWater Technology. The chip is composed of a driver RISC-V processor and a probabilistic design area, with connections to the ASIC control board as shown in Fig. 3c. The PIM design area implements 1143 p-bits within an area of 2.9 mm × 3.5 mm, of which approximately 3.11 mm$^2$ is utilized. The RISC-V processor provides control signals to the adjacent VCMA-TRNG board to generate and read random numbers before transmitting them to the design area. Experimental procedures and parameters are set by compiled C code which is executed on the processor.

In each clock cycle of the factorization circuit, the ASIC used up to 11 random bits, depending on the selected factorizer size, to select a single p-bit to be updated. The probabilistic logic unit (PLU), shown in Fig. 4a, used 16 random bits to implement a hyperbolic arc-tangent weighted random number via a LUT. This random number was compared to the input of a p-bit, shown in Fig. 4b, to determine its updated state. This version of the probabilistic computing update procedure is equivalent to Equation 1 but is easier to implement in hardware. Consequently, this circuit sampled from the energy distribution of the Ising Hamiltonian defined by the $J$ matrix and $h$ vector.

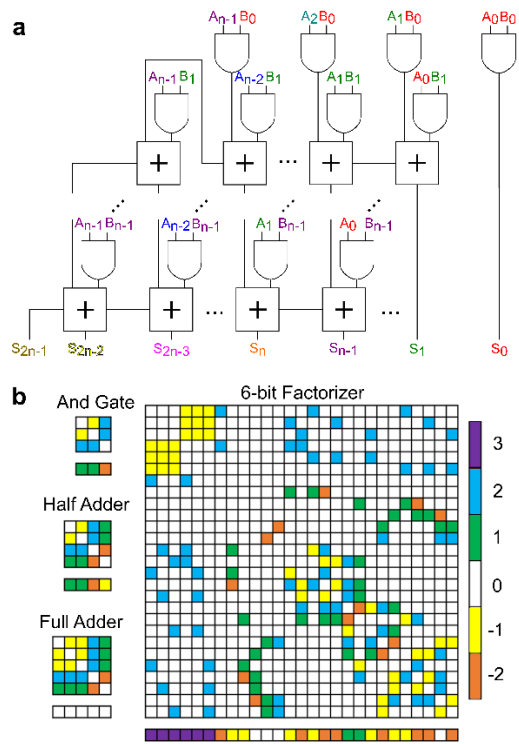

**Figure 5. Factorization circuit. a** Invertible logic gate factorization circuit made of invertible adders and gates. This design can be scaled to implement any symmetric binary factorization problem. **b** $J$ matrix (above) and $h$ vector (below) for an invertible AND gate, half adder, full adder, and 6-bit factorizer, respectively.

We used invertible logic gates[30,48], i.e., collections of p-bits whose equi-energy ground states correspond to the valid truth table states of the corresponding logic gates, as the basis of our factorizer design, following an approach proposed in our previous work[37]. These invertible gates are bi-directional: Either the inputs or the outputs can be fixed, resulting in logically consistent values appearing in the other[9,49]. Collections of AND gates, half adders, and full adders were connected into a standard multiplier design, shown in Fig. 5a. By fixing the output of this multiplier to a specified number, its factors are thus expected to appear on the invertible circuit's inputs once the p-bit network approaches its energy minimum. The p-bits within the PIM are grouped into designs of sizes 6 to 20 bits, each of which is defined by a static $J$ matrix and $h$ vector implemented at the circuit level. The $J$ matrix and $h$ vector for the 6-bit design, which utilizes 27 p-bits, are shown in Fig. 5b. Notably, the $J$ matrix and $h$ vector within each design do not need to be modified to factor different numbers, easing hardware implementation and making the circuit capable of factorizing arbitrary numbers up to a maximum set by the number of its p-bits. Moreover, the $J$ matrix is relatively sparse, which limits the possibility of routing issues stemming from the connectivity between p-bits.

The PIM ASIC has three logical connections to the VCMA-TRNG board, shown in Fig. 3a: The WRITE control signal (indicated in blue) turns ON a field-effect transistor (FET) on the VCMA-TRNG board, thereby allowing for voltage pulses from an external pulse generator to pass through to devices on the V-MTJ chip. The READ control signal (indicated in green) enables the readout mode of the VCMA-TRNG board, allowing for the resistance of the V-MTJ to be sensed. The OUT signal (indicated in yellow) signifies whether the V-MTJ is in the parallel or antiparallel state after a digitization using the readout circuit of Fig. 3d, which is also implemented directly on the VCMA-TRNG board. The readout circuit uses a voltage divider of the V-MTJ and a resistor, whose resistance is chosen to fall between the parallel and antiparallel resistances of the V-MTJ, alongside a comparator, to convert the V-MTJ state into a digitized voltage signal. As shown in Table E1, a bit stream of length 900,000, measured using the VCMA-TRNG board, passed the National Institute of Standards and Technology (NIST) Statistical Test Suite (STS)[65] after passing through two exclusive-OR (XOR4) layers. However, the bit streams supplied to the PIM ASIC were generated directly from the VCMA-TRNG and did not go through any XOR layers.

Each iteration, defined as a time-step in which one p-bit may change states, first involves the transfer of a list of up to 27 random bits, based on the problem size, from the VCMA-TRNG board to the RISC-V processor built into the ASIC. Once gathered, these bits are transferred to the p-bit area shown in Fig. 3c, where a number of them, depending on the factorizer size chosen, are then used to choose the p-bit to be updated during the iteration. In addition, 16 of these random bits are used to sample from the hyperbolic tangent distribution used in performing the update calculations. As a result, the selected p-bit is updated within a single ASIC clock cycle. At the end of each iteration, the ASIC adjusts the temperature in a linear fashion to implement a simulated annealing procedure. When a set number of iterations have been completed, the RISC-V processor reads, from the p-bit area, the final state of the p-bit circuit for storage in on-chip memory, before starting a new trial. Due to the probabilistic nature of the Ising machine, there is a finite chance of an invalid state being represented, necessitating multiple trials to reduce the uncertainty associated with the solution.

## V. Results and Discussion

A simulated parameter sweep, shown in Fig. 6a, using a series of 6-bit factorization problems was conducted to explore the temperature dependence of the PIM's behavior. It was found that the PIM's performance was much more sensitive to the final temperature compared to the starting temperature used in simulated annealing. A benefit of the used design is that in addition to performing factorization by fixing the invertible multiplier's output, multiplication can be performed by fixing its inputs, and division can be performed by fixing its output and one of its inputs. Using a pair of values within this optimal region for factorization, 8192 iteration-long simulated trials for multiplication of 7 with 5 and division of 35 by 5 were conducted.

The distribution of the final states of the PIM correctly corresponded to a product of 35 in a plurality of cases, shown in Fig. 6b, and a quotient of 7 in a majority of cases, shown in Fig. 6c, for each problem respectively.

With the same simulated annealing parameters, a series of 8192 iteration-long 6-bit factorization trials, shown in Fig. 6d, were conducted on the experimental setup. The two valid states, (5, 7) and (7, 5), both occurred at a much greater frequency than the invalid states, indicating that the system can successfully solve this factorization problem. The simulated energy evolution of the p-bits within the 6-bit factorization circuit used in one of these trials is shown in Fig 6e. As the temperature of the system decreases, the energy does as well, before the system settles into one of the two correct solutions. The derivations of the $J$ and $h$ matrices used for these demonstrations are discussed in Supplementary Note 2.

Additionally, we explored a slight modification of our original p-bit mapping design, which is not yet implemented in the current ASIC. By considering only the factorization of odd numbers, we were able to decrease the number of p-bits used at each factorization size. More crucially, we introduced the ability to have the evolution of the probabilistic system halt when a correct solution was found, by checking the validity of the current state using a normal multiplier design. This concept of running the forward circuit to check potential solutions created by the PIM can, in principle, be applied to many other optimization problems as well[40]. In this revised design a 20-bit factorizer is composed of 275 p-bits, whose interactions are defined by the $J$ matrix and $h$ vector shown in Fig. S4 (Supplementary Note 2). A frequency plot for simulations of an arbitrary 20-bit factorization problem in Fig. 6f shows convergence in a clear plurality of trials.

The size of the problems on which we were able to effectively gather statistics was limited by a few factors, the largest of which was the delay in driving the I/O ports of the ASIC control board that was used to communicate with the VCMA-TRNG PCB. Using the RISC-V CPU, it takes approximately 112 μs to drive the READ signal high and 106 μs to record the resultant random state. To ensure signal propagation through all components, a conservative 3.26 ms write/read procedure was used. With this procedure, up to 80 write pulses could be sent per random bit, limiting the lifespan of each wire-bonded V-MTJ to an observed approximate 10,000,000 bits. However, this I/O problem can, in principle, be addressed by (i) direct I/O integration with the ASIC, (ii) merging the VCMA-TRNG and ASIC control boards, and (iii) ultimately, by monolithic integration of the V-MTJs on the ASIC itself. While only a single V-MTJ was used in this work, such integration would shift the bottleneck in achievable iterations per second to that of the CMOS hardware if several V-MTJs were used to generate random bits concurrently. Notably, the 10 ns write time does not limit the maximum overall operation frequency, as the CMOS hardware can on different clock cycles pull from alternating groups of V-MTJs which are running on slower offset clocks. A comparison of the experimental and simulated factorizers to other MTJ-based PIMs is presented in Table 1.

To evaluate the ultimate scalability of this approach, the 130 nm ASIC design was subsequently adapted for the new p-bit mapping and implemented on two predictive PDKs for more advanced CMOS nodes. While not directly manufacturable, the Cadence 45 nm Generic PDK (GPDK) and the 7 nm Arizona State Predictive PDK (ASAP)[66] offer good approximations of the transistor densities of their respective nodes. These were used to implement a 40-bit factorizer and an 80-bit factorizer, respectively. The GPDK-defined 40-bit factorizer had a synthesized area of 0.120 mm$^2$ and a routed area of 0.176 mm$^2$. Greater improvements in area were seen by using the 7 nm ASAP to implement an 80-bit factorizer. Despite implementing a much more complicated problem, the design was only 0.0295 mm$^2$ synthesized and 0.0440 mm$^2$ when routed.

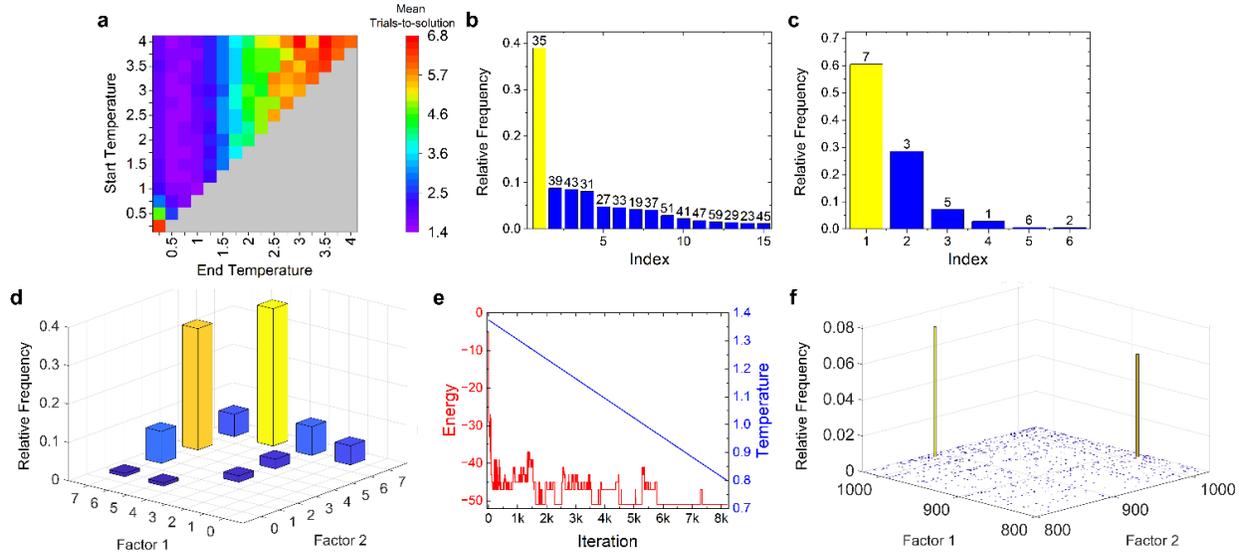

**Figure 6. PIM performance results. a** Simulated mean trials-to-solution for 6-bit factorization problem instances with different starting and finishing temperatures used in simulated annealing. Each trial consisted of 8192 iterations, after which the p-bit states were recorded. A relatively wide band of temperature parameters has equivalently good performance. **b** Frequency plot of the invertible multiplier's output bits when the input bits are fixed at values of 7 and 5, respectively. The correct product of 35 appears in a clear plurality of cases. The temperature was swept from 1.375 to 0.8 over 8192 trials. **c** Frequency plot of one of the invertible multiplier's inputs when the other is fixed at 5 and the output fixed at 35. The correct value of the quotient (i.e., 7) appears on the free input in a majority of trials using the same parameters as before. **d** Experimentally measured frequency plot for the 6-bit factorization problem of the number 35 after 8192 iterations, in which the temperature was swept from 1.375 to 0.8. Collectively, the two valid states make up 67.5% of the final results. **e** Plot showing the time evolution of the energy of the relevant p-bits used for a 6-bit factorization trial with the same parameters as above. **f** Frequency plot for the end state of simulated factorization attempts of the 20-bit semiprime 894,479 using the modified design described in Section V of the main text. Temperature was linearly swept between 20 and 4 with a timeout of $2^{18}$ iterations. The two valid solution pairs including 883 and 1013 appear most frequently.

The above examples indicate that relatively large Ising representations may be implemented by rather modest circuit areas in advanced nodes, due to the increased p-bit density. To investigate this possibility more abstractly, we further investigated the scaling trend by considering the reported densities of the used SkyWater 130 nm PDK and various Intel nodes, as seen in Fig. E1. In these cases, the area of the PLU and CMOS-based random number generator modules were ignored. This is a valid assumption for systems with serially updating p-bits, as their area remains relatively constant with the number of p-bits. For instance, the PLU and CMOS-based random number generator represent only around 9% of the 40-bit factorizer design and 2% of the 80-bit design area in the 45 nm and 7 nm designs discussed above. Our implemented designs line up closely with this projection, with the exception of the manufactured SkyWater 130 nm chip. This deviation can be explained by the low density target that was used for synthesis of this chip, which was chosen conservatively due to the larger-than-needed available design area. The co-integration of V-MTJs would allow further area reductions, as an equivalent V-MTJ true random number generator module would use an estimated < 0.1% of the area utilized for the corresponding CMOS-based one in both the 45 nm and 7 nm designs (See Methods for details of this estimation).

For circuits where p-bits are updated in parallel, both the V-MTJ random number generator modules and the PLU modules need to be duplicated to a degree dependent on the chosen architecture. As an example, we can split our p-bits into 275 groups, corresponding to our 20-bit revised factorizer design, where the p-bits in each group update in parallel (i.e., each p-bit on average updates after 275 cycles). In this case, the total area of our 45 nm design and 7 nm design would be an estimated 12% and 18% larger than that reported in Fig. E1a, respectively. These values are likely pessimistic because the distribution of the PLU and random number modules reduces the area needed to route their outputs throughout the circuit. Furthermore, the PLU's size can be significantly reduced by decreasing the precision of its hyperbolic arctangent implementation from its current 16-bit value. Nevertheless, this suggests that, for the 7 nm design, an approximately 17× increase in p-bit update rate can be achieved at the cost of an additional 18% area utilization compared to the serial case using a CMOS random number generator module. As with the estimate provided above, using a V-MTJ TRNG can significantly alleviate the added area due to additional random number generators in such a parallel-updating design.

These estimates illustrate the key advantages of ASICs for PIM implementation: The first is that the number of p-bits that can be implemented in a relatively small area is incredibly large for modern nodes. The second is that fast true random number generators based on V-MTJs, requiring only a single clock cycle to generate a random bit, can be implemented in close proximity to the logic. As seen from Fig. E1b, ASICs implemented in sub-10 nm nodes may in fact provide sufficient numbers of p-bits within a typical chip area to make significantly larger Ising representations accessible with this hardware approach. Note, however, that this projection only illustrates the achievable hardware sophistication at each node, and does not evaluate the time required to reach a solution. The latter is algorithm-dependent, and we expect that more sophisticated annealing algorithms will be increasingly required as the node size is scaled down. Moreover, while not explored in this work, using a larger number of V-MTJs could allow for multiple p-bits to be updated in parallel during a single clock cycle, thereby potentially offering time-to-solution improvements.

| Reference | CMOS Platform | Architecture | P-Bit Count | Digital-Analog Conversion | Entropy Source | MTJ Configuration | Bit Rate (kBit/s) | Energy per bit (pJ) | TMR Ratio | MTJ Size (nm) |
|---|---|---|---|---|---|---|---|---|---|---|
| This Work (Experiment) | ASIC | Synchronous | 1143 | No | V-MTJ | Out-of-Plane | 0.307 | 34.4 | 150% | 30 |
| This Work (Projection) | ASIC | Synchronous | 1143 | No | V-MTJ | Out-of-Plane | 10,000 | 0.43 | 150% | 30 |
| Shao et al [37] | CPU | Synchronous | 38 | No | V-MTJ | Out-of-Plane | - | 1.94 | 150% | 70 |
| Si et al [33] (Experiment) | MCU | Synchronous | 80 | Yes | S-MTJ | Out-of-Plane | 10 | - | 103% | 50 |
| Si et al [33] (Projection) | ASIC | Synchronous | 4096 | Yes | S-MTJ | In-Plane | 1,100 | - | 110% | 60 |
| Borders et al [6] | MCU | Asynchronous | 8 | Yes | S-MTJ | Out-of-Plane | 8 | 1027 | 100% | 40-80 |
| Singh et al [34] | FPGA | Asynchronous | 5 | No | S-MTJ | Out-of-Plane | 5 | 1691 | 65% | 50 |

**Table 1. CMOS+MTJ PIM comparison.** A comparison of the CMOS platform used, the system architecture, p-bit count, requirement of digital-to-analog conversion, MTJ type and magnetic configuration, bit rate, energy per random bit, TMR ratio, and MTJ diameter. Energy per bit values are provided where reported, or where they could be estimated from data. We define synchronous architectures as systems where p-bits are updated based on a clock, as opposed to asynchronous architectures where p-bit updates are triggered by a probabilistic clock enabled by the MTJ. Energy per bit calculations are based on the power consumption of the entropy source and the read/write speed for synchronous architectures, or the mean dwell time for asynchronous ones. Asynchronous architectures require larger energy per random bit arriving from the MTJ, due to the need to apply a continuous voltage to the device. We note, however, that the MTJ in these asynchronous circuits is typically used in conjunction with a conventional CMOS random number generator, and thus contributes fewer random bits to the same computing task. Bit rate calculations for asynchronous architectures are determined by the system's bottleneck, which is either the dwell time of the entropy source or the operating frequency. Synchronous bit rates are limited only by the system's clock speed, assuming random bits can be generated and read at a speed higher than the clock cycle. Unavailable quantities are left blank.

## Methods

The quality of generated true random numbers was evaluated using the National Institute of Standards and Technology (NIST) statistical testing suite[65]. A stream of 900,000 VCMA-generated random bits was collected from the VCMA-TRNG PCB. The data was divided into 100 equal groups, and all eleven NIST tests were run on each group. The data was considered statistically random if 98 out of 100 groups passed a single test. The initial bit stream only passed two of the tests, indicating that the data collected from the V-MTJ exhibited some bias towards one of the states. We attribute this to the residual stray field that was not fully compensated, as well as thermal drift during the measurement. This bias could be mitigated by performing an exclusive-OR (XOR) operation between bits in the bit stream. A two-input XOR gate returns a 0 state if the two inputs are the same, and a 1 if the inputs are different. Since the two input bits are randomly generated from the V-MTJ, the XOR operation helps shift the bit stream towards an equal distribution of 0 and 1 states[2]. However, performing an XOR operation requires sacrificing half of the bit stream's total length. The NIST tests were rerun on the XOR2 bit stream, now of length 450,000, demonstrating considerable improvement. By conducting another XOR operation on the already XORed data, i.e., XOR4 on the original data, all 11 NIST tests achieved at least a 99/100 success rate, deeming the bit stream truly random. Table E1 shows the NIST test results at each stage in the XORing process.

The probabilistic computing representation for the factorization problem was produced using self-made Python 3 code and simulated using self-made C++ code. Verilog representation of the ASIC was written, which was in turn used as the input for the OpenLane RTL to GDSII pipeline, which utilized the SkyWater 130 nm open-source process design kit (PDK). The ASIC was manufactured via the Efabless multi-project wafer (MPW) service, which incorporated an existing design for a small CPU based on a VexRiscv minimal+debug configuration[67]. An internal 10 MHz oscillator was used as the clock during verification. The two other designs were synthesized using the Cadence Genus software. They were then implemented using the Cadence Innovus Software with a core density target of 70%. Medium effort was used when applicable. For the 7 nm design, the Low Voltage Threshold (LVT) ASAP7 5-track cell library was used. While much larger designs could reasonably be implemented on both PDKs, we were limited to an 80-bit factorizer due to memory limitations.

For the scaling projection calculations, a random 5-connected p-bit from the manufactured design was synthesized, including all components calculating its input value, and the number of gates and transistors were counted. The number of gates was compared to the self-reported routed gate density of the SkyWater 130 nm High Density Standard Cell Library[68]. The number of transistors was compared to the densities reported by Intel at IEDM 2018[69], with labeling reflecting the renaming to the Intel 7 and Intel 4 node since then[70]. The V-MTJ area was estimated based on an assumed 50 $F^2$ cell size[71-73] and an additional equivalent area for support circuitry. The size of a V-MTJ module was calculated as this value multiplied by the sum of 16 and the number of addressing bits. The parallel area projections were calculated by taking the original area, subtracting the area of the original CMOS random number generator and PLU, and then adding the area sum of a V-MTJ module and PLU multiplied by the number of p-bit groups.

During operation, a small in-plane field was applied to the V-MTJs via a permanent magnet suspended above the experimental setup. A positioner was used to change the magnet's height and angle to produce a bias field permitting 50%/50% VCMA switching probability. A pulse generator with voltage increments of approximately 100 mV was used, and thus it is possible that identical switching can be performed with a lesser voltage. Similarly, pulse widths of less than 10 ns were not tested. It is worth noting that the permanent bias field can in principle be integrated within the device, by adding a fixed magnetic layer that is coupled with the free layer through dipole interaction, as shown in previous work[74].

## Author contributions

PKA initiated the project and conceived the idea with contributions from YS, ER, CD, JA, GF and KYC. CD and JA designed the ASIC chip, implemented the TRNG board, and performed related testing and simulations. PKA designed the V-MTJ devices with contributions from YS. NDM and JAK fabricated the V-MTJ devices. CD, JA, and PKA wrote the manuscript with contributions from the other authors. All authors discussed the results, contributed to the data analysis, and commented on the manuscript. CD and JA contributed equally to this work. The study was performed under the leadership of PKA.

## Acknowledgements

This work was supported by the U.S. National Science Foundation (NSF) under award numbers 2322572, 2425538, 2311296, and 2400463, and by a gift from Nokia corporation. The authors thank Canon ANELVA corporation for part of the magnetic thin film deposition and characterization. E.R. and G.F. acknowledge financial support by the project PRIN 2020LWPKH7 funded by the Italian Ministry of Universities and Research (MUR), and the PETASPIN association (www.petaspin.com). E.R. also acknowledges the project PON Capitale Umano (CIR_00030) funded by the Italian Ministry of Universities and Research (MUR).

## Data availability statement

The data that support the findings of this study are available upon reasonable request from the corresponding author.

## Competing interests

The authors declare no competing interests.

## References

1. Finocchio, G. *et al.* The promise of spintronics for unconventional computing. *J. Magn. Magn. Mater.* 521, 167506 (2021).
2. Zhang, P., Peeta, S. & Friesz, T. Dynamic game theoretic model of multi-layer infrastructure networks. *Networks Spatial Econ.* 5, 147–178 (2005).
3. Schuetz, M. J., Brubaker, J. K. & Katzgraber, H. G. Combinatorial optimization with physics-inspired graph neural networks. *Nat. Mach. Intell.* 4, 367–377 (2022).
4. Finocchio, G. *et al.* Roadmap for unconventional computing with nanotechnology. *Nano Futur.* (2023).

5. Kalinin, K. P. & Berloff, N. G. Large-scale sustainable search on unconventional computing hardware. *arXiv preprint arXiv:2104.02553* (2021).

6. Borders, W. A. et al. Integer factorization using stochastic magnetic tunnel junctions. *Nature* 573, 390–393 (2019).

7. Mohseni, N., McMahon, P. L. & Byrnes, T. Ising machines as hardware solvers of combinatorial optimization problems. *Nat. Rev. Phys.* 4, 363–379 (2022).

8. Tanahashi, K., Takayanagi, S., Motohashi, T. & Tanaka, S. Application of Ising machines and a software development for Ising machines. *J. Phys. Soc. Jpn.* 88, 061010 (2019).

9. Aadit, N. A. *et al.* Massively parallel probabilistic computing with sparse Ising machines. *Nat. Electron.* 5, 460–468 (2022).

10. Newell, G. F. & Montroll, E. W. On the theory of the Ising model of ferromagnetism. *Rev. Mod. Phys.* 25, 353 (1953).

11. Griffiths, R. B. Correlations in Ising ferromagnets. i. *J. Math. Phys.* 8, 478–483 (1967).

12. Goto, H., Lin, Z. & Nakamura, Y. Boltzmann sampling from the Ising model using quantum heating of coupled nonlinear oscillators. *Sci. reports* 8, 7154 (2018).

13. Kanao, T. & Goto, H. High-accuracy Ising machine using Kerr-nonlinear parametric oscillators with local four-body interactions. *npj Quantum Inf.* 7, 18 (2021).

14. Marandi, A., Wang, Z., Takata, K., Byer, R. L. & Yamamoto, Y. Network of time-multiplexed optical parametric oscillators as a coherent Ising machine. *Nat. Photonics* 8, 937–942 (2014).

15. Okawachi, Y. *et al.* Demonstration of chip-based coupled degenerate optical parametric oscillators for realizing a nanophotonic spin-glass. *Nat. communications* 11, 4119 (2020).

16. Okawachi, Y. *et al.* Nanophotonic spin-glass for realization of a coherent Ising machine. *arXiv preprint arXiv:2003.11583* (2020).

17. Tezak, N. *et al.* Integrated coherent Ising machines based on self-phase modulation in microring resonators. *IEEE J. Sel. Top. Quantum Electron.* 26, 1–15 (2019).

18. Lo, H., Moy, W., Yu, H., Sapatnekar, S. & Kim, C. H. An ising solver chip based on coupled ring oscillators with a 48-node all-to-all connected array architecture. *Nat. Electron.* 6, 771–778 (2023).

19. Karpuzcu, U. *et al.* Cobi: A coupled oscillator based Ising chip for combinatorial optimization. *Nat. Portfolio: Manuscr. (2024).*

20. Bashar, M. K., Mallick, A. & Shukla, N. Experimental investigation of the dynamics of coupled oscillators as Ising machines. *IEEE Access* 9, 148184–148190 (2021).

21. Mallick, A. *et al.* Using synchronized oscillators to compute the maximum independent set. *Nat. communications* 11, 4689 (2020).

22. Bashar, M. K., Lin, Z. & Shukla, N. Stability of oscillator Ising machines: Not all solutions are created equal. *J. Appl. Phys.* 134 (2023).

23. Bashar, M. K., Li, Z., Narayanan, V. & Shukla, N. An FPGA-based max-k-cut accelerator exploiting oscillator synchronization model. In *2024 25th International Symposium on Quality Electronic Design (ISQED)*, 1–8 (IEEE, 2024).

24. Tatsumura, K. *et al*. FPGA-based simulated bifurcation machine. *FPL* 29 (2019).

25. Tatsumura, K. *et al.* Scaling out Ising machines using a multi-chip architecture for simulated bifurcation. *Nat. Electron.* 4, 3 (2021).

26. Kashimata, T. *et al.* Efficient and scalable architecture for multiple-chip implementation of simulated bifurcation machines. *IEEE Access* (2024).

27. Litvinenko, A. *et al.* A spinwave Ising machine. *Commun. Phys.* 6, 227 (2023).

28. González, V., Litvinenko, A., Khymyn, R. & Åkerman, J. Global biasing using a hardware-based artificial Zeeman term in spinwave Ising machines. In *2023 IEEE International Magnetic Conference-Short Papers (INTERMAG Short Papers)*, 1–2 (IEEE, 2023).

29. Litvinenko, A., Khymyn, R., Ovcharov, R. & Åkerman, J. A 50-spin surface acoustic wave ising machine. *arXiv preprint arXiv:2311.06830* (2023).

30. Camsari, K. Y., Sutton, B. M. & Datta, S. P-bits for probabilistic spin logic. *Appl. Phys. Rev.* 6 (2019).

31. Kaiser, J. & Datta, S. Probabilistic computing with p-bits. *Appl. Phys. Lett.* 119 (2021).

32. Grimaldi, A. *Evaluating spintronics-compatible implementations of Ising machines*. *Phys. Rev. Appl.* 20, 024005 (2023).

33. Si, J. *et al.* Energy-efficient superparamagnetic ising machine and its application to traveling salesman problems. *Nat. Commun.* 15, 3457 (2024).

34. Singh, N. S. *et al.* CMOS plus stochastic nanomagnets enabling heterogeneous computers for probabilistic inference and learning. *Nat. Commun.* 15, 2685 (2024).

35. Hamerly, R. *et al.* Experimental investigation of performance differences between coherent ising machines and a quantum annealer. *Sci. advances* 5, eaau0823 (2019).

36. Weinberg, P. *et al.* Scaling and diabatic effects in quantum annealing with a d-wave device. *Phys. Rev. Lett.* 124, 090502 (2020).

37. Shao, Y. *et al.* Probabilistic computing with voltage-controlled dynamics in magnetic tunnel junctions. *Nanotechnology* 34, 495203 (2023).

38. Aadit, N. A. *et al.* Computing with invertible logic: Combinatorial optimization with probabilistic bits. In *2021 IEEE International Electron Devices Meeting (IEDM)*, 40–3 (IEEE, 2021).

39. Imam, R., Areeb, Q. M., Alturki, A. & Anwer, F. Systematic and critical review of rsa based public key cryptographic schemes: Past and present status. *IEEE Access* 9, 155949–155976 (2021).

40. Grimaldi, A. *et al.* Spintronics-compatible approach to solving maximum-satisfiability problems with probabilistic computing, invertible logic, and parallel tempering. *Phys. Rev. Appl.* 17, 024052 (2022).

41. Chowdhury, S., Camsari, K. Y. & Datta, S. Accelerated quantum monte carlo with probabilistic computers. *Commun. Phys.* 6, 85 (2023).

42. Kuon, I. & Rose, J. Measuring the gap between FPGAs and ASICs. In *Proceedings of the 2006 ACM/SIGDA 14th international symposium on Field programmable gate arrays*, 21–30 (2006).

43. Fukushima, A. *et al.* Recent progress in random number generator using voltage pulse-induced switching of nano-magnet: A perspective. *APL Mater.* 9.3 (2021).

44. Shao, Y. & Khalili Amiri, P. Progress and application perspectives of voltage-controlled magnetic tunnel junctions. *Adv. Mater. Technol.* 8, 2300676 (2023).

45. Borders, W. A. *et al.* Integer factorization using stochastic magnetic tunnel junctions. *Nature* 573, 390–393 (2019).

46. Sutton, B., Camsari, K. Y., Behin-Aein, B. & Datta, S. Intrinsic optimization using stochastic nanomagnets. *Sci. reports* 7, 44370 (2017).

47. Camsari, K. Y. *et al.* From charge to spin and spin to charge: Stochastic magnets for probabilistic switching. *Proc. IEEE* 108, 1322–1337 (2020).

48. Lv, Y., Bloom, R. P. & Wang, J.-P. Experimental demonstration of probabilistic spin logic by magnetic tunnel junctions. *IEEE Magn. Lett.* 10, 1–5 (2019).

49. Camsari, K. Y., Faria, R., Sutton, B. M. & Datta, S. Stochastic p-bits for invertible logic. *Phys. Rev. X* 7, 031014 (2017).

50. Camsari, K. Y., Salahuddin, S. & Datta, S. Implementing p-bits with embedded MTJ. *IEEE Electron Device Lett.* 38, 1767–1770 (2017).

51. Ikeda, S. *et al.* A perpendicular-anisotropy CoFeB–MgO magnetic tunnel junction. *Nat. materials* 9, 721–724 (2010).

52. Worledge, D. *et al.* Spin torque switching of perpendicular Ta|CoFeB|MgO-based magnetic tunnel junctions. *Appl. physics letters* 98 (2011).

53. Khalili Amiri, P. *et al.* Switching current reduction using perpendicular anisotropy in CoFeB–MgO magnetic tunnel junctions. *Appl. Phys. Lett.* 98 (2011).

54. Fukushima, A. *et al.* Spin dice: A scalable truly random number generator based on spintronics. *Appl. Phys. Express* 7, 083001 (2014).

55. Jia, X. *et al.* Spintronics based stochastic computing for efficient bayesian inference system. *Proc. 23rd Asia and South Pacific Design Automation Conference (ASP-DAC)*, 580–585 (IEEE, 2018).

56. Zink, B. R., Lv, Y. & Wang, J.-P. Review of magnetic tunnel junctions for stochastic computing. *IEEE J. on Explor. Solid-State Comput. Devices Circuits* 8, 173–184 (2022).

57. Liu, S. *et al.* Random bitstream generation using voltage-controlled magnetic anisotropy and spin orbit torque magnetic tunnel junctions. *IEEE J. on Explor. Solid-State Comput. Devices Circuits* 8, 194–202 (2022).

58. Maruyama, T. *et al.* Large voltage-induced magnetic anisotropy change in a few atomic layers of iron. *Nat. nanotechnology* 4, 158–161 (2009).

59. Amiri, P. K. & Wang, K. L. Voltage-controlled magnetic anisotropy in spintronic devices. In *Spin*, vol. 2, 1240002 (World Scientific, 2012).

60. Shao, Y. *et al.* Sub-volt switching of nanoscale voltage-controlled perpendicular magnetic tunnel junctions. *Commun. Mater.* 3, 87 (2022).

61. Aggarwal, S. *et al.* Demonstration of a reliable 1 Gb standalone spin-transfer torque mram for industrial applications. In *2019 IEEE International Electron Devices Meeting (IEDM)*, 2–1 (IEEE, 2019).

62. Spin-transfer torque MRAM technology. https://www.everspin.com/spin-transfer-torque-mram-technology. Accessed: 2024-02-24.

63. Parkin, S., More, N. & Roche, K. Oscillations in exchange coupling and magnetoresistance in metallic superlattice structures: Co/ru, co/cr, and fe/cr. *Phys. review letters* 64, 2304 (1990).

64. Duine, R., Lee, K.-J., Parkin, S. S. & Stiles, M. D. Synthetic antiferromagnetic spintronics. *Nat. physics* 14, 217–219 (2018).

65. Rukhin, A. *et al.* A statistical test suite for random and pseudorandom number generators for cryptographic applications. Vol. 22. Gaithersburg, MD, USA: *US Department of Commerce, Technology Administration, National Institute of Standards and Technology* (2001).

66. Clark, L. T. *et al.* Asap7: A 7-nm finfet predictive process design kit. *Microelectron. J.* 53, 105–115 (2016).

67. Caravel management soc - litex. https://caravel-mgmt-soc-litex.readthedocs.io/en/latest/ (2022). Accessed: 2024-06-25.

68. Skywater foundry provided standard cell libraries. https://skywater-pdk.readthedocs.io/en/main/contents/libraries/ foundry-provided.html. Accessed: 2024-02-24.

69. Intel's 10nm Cannon Lake and core i3-8121u deep dive review. https://www.anandtech.com/show/13405/ intel-10nm-cannon-lake-and-core-i3-8121u-deep-dive-review/3 (2019). Accessed: 2024-05-17.

70. Intel process roadmap through 2025: Renamed process nodes, angstrom era begins. https://www.tomshardware.com/news/ intel-process-packaging-roadmap-2025 (2021). Accessed: 2024-05-17.

71. Lee, H., *et al*. Array-level analysis of magneto-electric random-access memory for high-performance embedded applications. *IEEE Magnetics Letters* 8, 1-5 (2017).

72. Sakimura, N., *et al*. MRAM cell technology for over 500-MHz SoC. *IEEE journal of solid-state circuits* 42.4, 830-838 (2007).

73. Lee, K., *et al*. 1Gbit high density embedded STT-MRAM in 28nm FDSOI technology. *2019 IEEE International Electron Devices Meeting (IEDM)* (2019).

74. Krizakova, V., *et al*. Field-free switching of magnetic tunnel junctions driven by spin–orbit torques at sub-ns timescales. *Applied Physics Letters* 116, 232406 (2020).

# Extended Data Figures

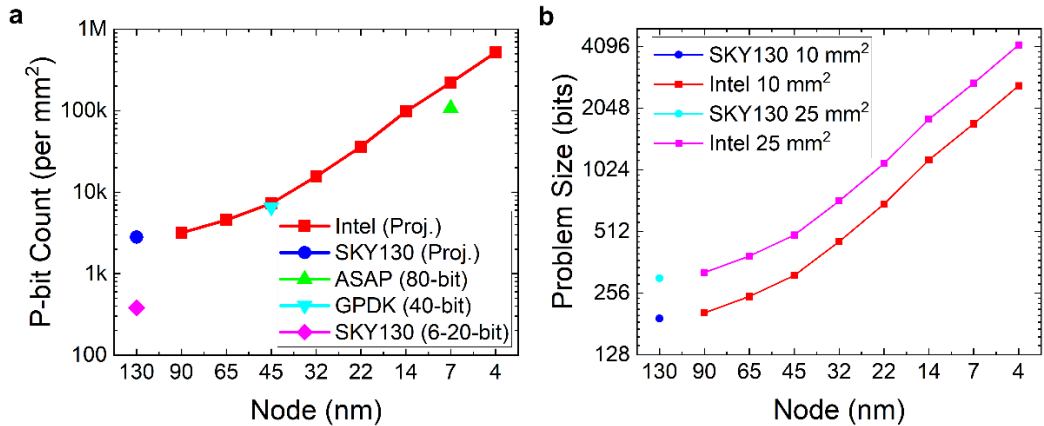

**Figure E1. Scaling projections. a** Approximate scaling of the count of p-bits that can be integrated per 1 mm$^2$ area at various node sizes. As the number of p-bits increases, their proportion of the total area approaches unity, so the area of other components can be neglected in the large number of p-bits limit. This is done for the Intel and SkyWater (SKY130) p-bit projection, in which self-reported transistor and routed gate densities are compared to synthesized designs for a random 5-connected p-bit from the factorization ASIC. The SKY130 6-20-bit factorizer plot represents the density of the manufactured design, which is low due to lack of tight density constraints applied during synthesis. The GPDK and ASAP points are routed designs created using Cadence Genus and Innovus. **b** Projected problem size (number of bits) that can be implemented at various node sizes, considering the SkyWater and Intel reported densities for a 10 mm$^2$ and 25 mm$^2$ design area, respectively.

| Test Name / XOR Stage | BASE | XOR2 | XOR4 |
|---|---|---|---|
| Frequency (Monobit) | 7/100 | 65/100 | 99/100 |
| Frequency within a Block | 21/100 | 85/100 | 99/100 |
| Cumulative Sums | 6/100 | 65/100 | 99/100 |
| Runs | 17/100 | 81/100 | 100/100 |
| Longest Run of Ones in Block | 28/100 | 83/100 | 99/100 |
| Binary Matrix Rank | 100/100 | 100/100 | 100/100 |
| Discrete Fourier Transform (Spectral) | 57/100 | 100/100 | 100/100 |
| Non-Overlapping Template Matching | 98/100 | 100/100 | 100/100 |
| Overlapping Template Matching | 57/100 | 100/100 | 99/100 |
| Maurer's Universal Statistical | 0/100 | 100/100 | 100/100 |
| Serial | 88/100 | 99/100 | 100/100 |

**Table E1.** Results of NIST statistical tests for randomness and the VCMA-TRNG PCB performance. Each test was run on a bit stream of 900,000 bits generated from the V-MTJ-based TRNG circuit. Each test splits the bit stream into 100 equal chunks and performs the test on each chunk to determine success or failure. NIST deems 98/100 passes or greater as a success. The XOR2 case required sacrificing half of the bit stream to skew the results closer to an equal distribution. The XOR4 case sacrificed three-quarters of the generated bits, but yielded results that passed all the test cases.